# Modelling and Analysis of the Distributed Coordination Function of IEEE 802.11 with Multirate Capability

F. Daneshgaran, M. Laddomada, F. Mesiti, and M. Mondin

*Abstract*— The aim of this paper is twofold. On one hand, it presents a multi-dimensional Markovian state transition model characterizing the behavior at the Medium Access Control (MAC) layer by including transmission states that account for packet transmission failures due to errors caused by propagation through the channel, along with a state characterizing the system when there are no packets to be transmitted in the queue of a station (to model non-saturated traffic conditions). On the other hand, it provides a throughput analysis of the IEEE 802.11 protocol at the data link layer in both saturated and non-saturated traffic conditions taking into account the impact of both transmission channel and multirate transmission in Rayleigh fading environment. Simulation results closely match the theoretical derivations confirming the effectiveness of the proposed model.

## I. INTRODUCTION

The IEEE802.11 MAC [1] presents a mandatory option, namely the Distributed Coordination Function (DCF), a medium access mechanism based on the CSMA/CA access method, that has received considerably attention in the past years [2]-[14].

A number of papers [3]-[5], after the seminal work by Bianchi, have addressed the problem of modelling the DCF in a variety of traffic load and channel transmission conditions. Most of them focuses on a scenario presenting $N$ saturated stations that transmit toward a common access point (AP) under the hypotheses that the packet rates along with the probability of transmission in a randomly chosen slot time is common to all the involved stations, while error events on the transmitted packets are mainly due to collisions between packets belonging to different stations.

Modeling of the DCF of IEEE 802.11 WLANs in unsaturated traffic conditions has been analyzed in a number of papers [6]-[10]. In [6] the authors extended the underlying model in order to consider non-saturated traffic conditions by introducing a new state, not present in the original Bianchi's model, accounting for the case in which the station queue is empty after successful completion of a packet transmission. Paper [7] proposes an extension of the Bianchi's model considering a new state for each backoff stage accounting for the absence of new packets to be transmitted, i.e., in unloaded traffic conditions.

This work was supported through funds provided by PRIN-ICONA project.
F. Daneshgaran is with ECE Dept., California State University, Los Angeles, USA.
M. Laddomada, F. Mesiti, and M. Mondin are with DELEN, Politecnico di Torino, Italy.

In [11], the authors look at the impact of channel induced errors and the received SNR on the achievable throughput in a system with rate adaptation whereby the transmission rate of the terminal is adapted based on either direct or indirect measurements of the link quality. In [12], authors observed that in multirate networks the aggregate throughput is strongly influenced by the bit rate of the slowest contending station: such a phenomenon is termed performance anomaly of the DCF of the IEEE 802.11 protocol. In [14], authors provide DCF models for finite load sources with multirate capabilities, while in [13] authors propose a DCF model for multirate networks and derive the saturation throughput. In both previous works, packet errors are only due to collisions between different contending stations.

In this paper, we substantially extend a previous work proposed in the companion papers [8]-[9] considering real channel conditions, both saturated and non-saturated traffic, and multirate capabilities. As a reference standard, we use network parameters belonging to the IEEE802.11b protocol, even though the proposed mathematical model holds for any flavor of the IEEE802.11 family or other wireless protocols with similar MAC layer functionality.

This paper is organized as follows. After a brief review of the functionalities of the contention window procedure at MAC layer, section II substantially extends the Markov model initially proposed by Bianchi, presenting modifications that account for transmission errors. Section III provides an expression for the aggregate throughput of the link, while Section IV derives the time slot duration needed for throughput evaluation. The adopted traffic model is discussed in Section V. Section VI briefly addresses the modelling of the physical layer of IEEE 802.11b in a variety of channel fading conditions. In section VII we present simulation results where typical MAC layer parameters for IEEE802.11b are used to obtain the throughput as a function of various system level parameters, and the SNR under typical traffic conditions.

## II. MARKOVIAN MODEL

In a previous paper [8], we proposed a bi-dimensional Markov model for characterizing the behavior of the DCF under a variety of real traffic conditions, both non-saturated and saturated traffic load, with packet queues of small sizes, and considered the IEEE 802.11b protocol with the basic 2-way handshaking mechanism. Many of the basic hypotheses are the same as the ones adopted by Bianchi in the seminal paper [2].

As a starting point for the derivations which follow, we adopt the bi-dimensional model proposed in the companion paper [8], appropriately modified in order to account for a scenario of $N$ contending stations each one employing a specific bit rate and a different transmission packet rate. For conciseness, we invite the interested reader to refer to [8]-[9] for many details on the considered bi-dimensional Markov model.

Consider the following scenario: $N$ stations transmit toward a common AP, whereby each station, characterized by an own traffic load, can access the channel using a data rate in the set $\{1, 2, 5.5, 11\}$ Mbps depending on channel conditions. Any bit rate is associated with a different modulation format, whereas the basic rate is 1 Mbps with DBPSK modulation (2Mbps with DQPSK if short preamble is used) [1]. We identify a generic station with the index $s \in S = \{1, 2, \cdots, N\}$, where $N$ is the number of stations in the network, and $S$ is the set of station indexes. As far as the transmission data rate is concerned, we define four rate-classes identified by a *rate-class identifier* $r$ taking values in the set $R = \{1, 2, 3, 4\}$ ordered by increasing data rates $R_D = \{1, 2, 5.5, 11\}$ Mbps (as an example, rate-class $r = 3$ is related to the bit rate 5.5 Mbps). Concerning control packets and PLCP header transmissions, the basic rate is identified by $R_C$.

The traffic load of the s-th station is identified by a packet arrival rate (PAR) $\lambda^{(s)}$ evaluated in packets per second. Upon defining both rate-classes and traffic, we can associate a generic station $s$ with a rate-class $r \in R$ and a proper traffic load $\lambda^{(s)}$. Therefore, we need to specify, with respect to the model proposed in [8], specific probabilities along with different Markov chains for each contending station in the network.

The two sources of errors on the transmitted packets are both collisions between packets and channel induced errors. In relation to the $s$-th station in the network, collisions can occur with probability $P_{col}^{(s)}$, while transmission errors due to imperfect channel can occur with probability $P_e^{(s)}$. Notice that $P_e^{(s)}$ depends upon the station rate-class $r$, which is in turn related to the received Signal-To-Noise (SNR) (appropriate expressions will be provided in Section VI for each rate-class). We assume that collisions and transmission error events are statistically independent. In this scenario, a packet from the $s$-th station is successfully transmitted if there is no collision (this event has probability $1 - P_{col}^{(s)}$) and the packet encounters no channel errors during transmission (this event has probability $1 - P_e^{(s)}$). The probability of successful transmission is therefore equal to $(1 - P_e^{(s)})(1 - P_{col}^{(s)})$, while the equivalent probability of failed transmission is defined as

$$P_{eq}^{(s)} = P_{col}^{(s)} + P_e^{(s)} - P_e^{(s)} \cdot P_{col}^{(s)} \quad (1)$$

To simplify the analysis, we make the assumption that the impact of channel induced errors on the packet headers are negligible because of their short length with respect to the data payload size [8].

The modified Markov model related to the $s$-th contending station is depicted in Fig. 1. We consider $(m+1)$ different backoff stages including the zero-th stage. The maximum contention windows (CW) size is $W_{max} = 2^m W_0$, and the notation $W_i = 2^i W_0$ is used to define the $i^{th}$ contention window size ($W_0$ is the minimum contention window size). A packet transmission is attempted only in the $(i, 0)$ states, $\forall i = 0, \ldots, m$. In case of collision, or due to the fact that transmission is unsuccessful because of channel errors, the backoff stage is incremented, so that the new state can be $(i+1, k)$ with uniform probability $P_{eq}^{(s)}/W_{i+1}$. The contention window is supposed to be common to all the stations in the network; for this reason the station index $s$ is dropped from the contention model depicted in Fig. 1.

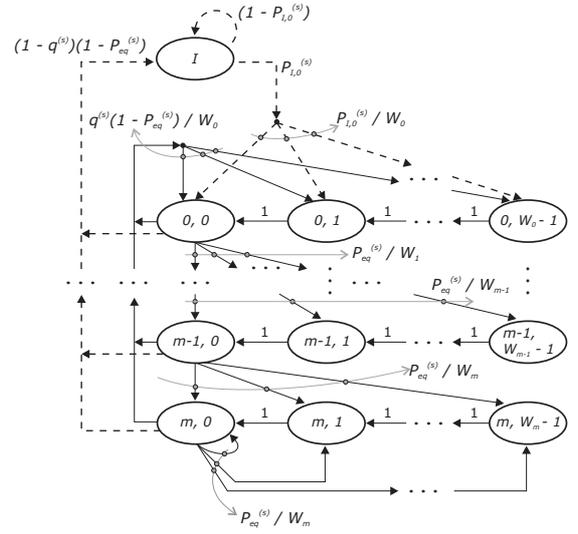

Fig. 1. Markov chain for the contention model of a generic station $s$ in general traffic conditions, based on the 2-way handshaking technique, considering the effects of channel induced errors.

In order to account for non-saturated traffic conditions, we introduced a new state labelled $I$, for the following two situations:

- Immediately after a successful transmission, the queue of the transmitting station is empty. This event occurs with probability $(1 - q^{(s)})(1 - P_{eq}^{(s)})$, whereby $q^{(s)}$ is the probability that there is at least one packet in the queue after a successful transmission,
- The station is in the idle state with an empty queue until a new packet arrival in the queue. Probability $P_{I,0}^{(s)}$ represents the probability that while the station resides in the idle state $I$ there is at least one packet arrival, and a new backoff procedure is scheduled.

We notice that the probability $P_{I,0}$ of residing in the idle state is strictly related to the adopted traffic model.

### III. MARKOVIAN PROCESS ANALYSIS

This section focuses on the evaluation of the stationary state distribution of the Markov model proposed in the previous section. The objective is to find the probability that a station occupies a given state at any discrete time slot along with the stationary probability $b_I^{(s)}$ of being in the idle state. This mathematical derivation is at the basis for the derivation of the probability $\tau^{(s)}$ that a station will attempt transmission in a randomly chosen slot time. For the sake of simplifying the notation, in what follows we will omit the apex $(s)$ since the mathematical derivations are valid for any contending station $s = 1, \ldots, N$. For future developments, from the model depicted in Fig. 1 we note the following relations:

$$\begin{array}{rcll} b_{i,0} &=& P_{eq} \cdot b_{i-1,0} = P_{eq}^i \cdot b_{0,0}, & \forall i \in [1, m-1] \\ b_{m,0} &=& \frac{P_{eq}^m}{1-P_{eq}} \cdot b_{0,0}, & i = m \end{array} \quad (2)$$

whereby $b_{i,j}$ is the stationary probability to be in the state labelled $i, j$ of the Markov chain in Fig. 1.

Let us focus on the meaning of the idle state $I$ noted in Fig. 1 to which the stationary probability $b_I$ is associated. It considers both the situation in which after a successful transmission there are no packets to be transmitted in the station queue, and the situation in which the packet queue is empty and the station is waiting for new packet arrivals. The stationary probability of being in state $b_I$ can be evaluated as

$$\begin{array}{rcl} b_I &=& (1-q)(1-P_{eq}) \sum_{i=0}^{m} b_{i,0} + (1-P_{I,0}) b_I \\ &=& \frac{(1-q)(1-P_{eq})}{P_{I,0}} \cdot \sum_{i=0}^{m} b_{i,0} \end{array} \quad (3)$$

Upon employing the probabilities $b_{i,0}$ noted in (2), it is straightforward to obtain:

$$\sum_{i=0}^{m} b_{i,0} = b_{0,0} \left[ \sum_{i=0}^{m-1} P_{eq}^i + \frac{P_{eq}^m}{1-P_{eq}} \right] = \frac{b_{0,0}}{1-P_{eq}} \quad (4)$$

By using the previous result, (3) simplifies to

$$b_I = \frac{1-q}{P_{I,0}} \cdot b_{0,0} \quad (5)$$

The other stationary probabilities for any $k \in [1, W_i-1]$ follow by resorting to the state transition diagram shown in Fig. 1:

$$b_{i,k} = \frac{W_i - k}{W_i} \begin{cases} q(1-P_{eq}) \cdot \sum_{i=0}^{m} b_{i,0} + \\ +P_{I,0} \cdot b_I, & i = 0 \\ P_{eq} \cdot b_{i-1,0}, & i \in [1, m-1] \\ P_{eq}(b_{m-1,0} + b_{m,0}), & i = m \end{cases} \quad (6)$$

Employing the normalization condition, after some mathematical manipulations, and remembering (4), it is possible to obtain:

$$1 = \sum_{i=0}^{m} \sum_{k=0}^{W_i-1} b_{i,k} + b_I = \alpha \cdot b_{0,0} + b_I \quad (7)$$

whereby

$$\alpha = \frac{1}{2} \left\{ W_0 \left[ \frac{1 - (2P_{eq})^m}{1 - 2P_{eq}} + \frac{(2P_{eq})^m}{1-P_{eq}} \right] + \frac{1}{1-P_{eq}} \right\} \quad (8)$$

From (7), the following equation for computation of $b_{0,0}$ easily follows:

$$b_{0,0} = \frac{1-b_I}{\alpha} \quad (9)$$

Equ. (9) is used to compute $\tau^{(s)}$, the probability that the $s$-th station starts a transmission in a randomly chosen time slot. In fact, taking into account that a packet transmission occurs when the backoff counter reaches zero, we have:

$$\begin{array}{rcl} \tau^{(s)} &=& \sum_{i=0}^{m} b_{i,0}^{(s)} = \frac{b_{0,0}^{(s)}}{1-P_{eq}^{(s)}} = \frac{1-b_I^{(s)}}{\alpha^{(s)}(1-P_{eq}^{(s)})} = \\ &=& \frac{2(1-b_I^{(s)})(1-2P_{eq}^{(s)})}{(W_0+1)(1-2P_{eq}^{(s)}) + W_0 P_{eq}^{(s)}(1-(2P_{eq}^{(s)})^m)} \end{array} \quad (10)$$

whereby we re-introduced the apex $(s)$ since this expression will be used in the following.

The collision probability $P_{col}^{(s)}$ needed to compute $\tau^{(s)}$ can be found considering that using a 2-way hand-shaking mechanism, a packet from a transmitting station encounters a collision if in a given time slot, at least one of the remaining $(N-1)$ stations transmits simultaneously one packet. Since each station has its own $\tau^{(s)}$, the collision probability for the $s$-th contending station depends on the transmission probabilities of the remaining stations as follows:

$$P_{col}^{(s)} = 1 - \prod_{\substack{j=1 \\ j \neq s}}^{N} (1 - \tau^{(j)}) \quad (11)$$

Given the set of $N$ equations (1) and (10), a non-linear system of $2N$ equations can be solved in order to determine the values of $\tau^{(s)}$ and $P_{col}^{(s)}$ for any $s = 1, \ldots, N$: this is the operating point corresponding to the $N$ stations in the network, needed in order to determine the aggregate throughput of the network, defined as the fraction of time the channel is used to successfully transmit payload bits:

$$S = \sum_{s=1}^{N} \frac{1}{T_{av}} P_s^{(s)} \cdot (1 - P_e^{(s)}) \cdot PL \quad (12)$$

whereby the summation is over the throughput related to the $N$ contending stations, $PL$ is the average payload size, and $T_{av}$ is the expected time per slot defined in the following.

Probabilities involved in (12) are as follows: $P_e^{(s)}$ is the PER (or FER) of the $s$-th station due to imperfect channel transmissions, and $P_s^{(s)}$ is the probability that a packet transmission from the $s$-th station is successful. In the next section, we derive the mathematical relations defining both $T_{av}$ and the probabilities involved in (12).

## IV. ESTIMATING THE AVERAGE TIME SLOT DURATION

In order to proceed further, we need to evaluate the average time $T_{av}$ spent by a station in any possible state. The average duration $T_{av}$ of a time slot can be evaluated by weighting the times spent by a station in a particular state with the probability of being in that state. It is possible to note four kind of time slots. The average idle slot duration, identified by $T_I$, in which no station is transmitting over the channel. The average collision slot duration, identified by $T_C$, in which more than one station is attempting to access the channel. The average duration of the slot due to erroneous transmissions because of imperfect channel conditions, identified by $T_E$. The average slot duration of a successful transmission, identified by $T_S$.

**The average idle slot duration.** The average idle slot duration can be evaluated as the probability $(1 - P_t)$ that no station is attempting to gain the access to the channel times the duration $\sigma$ of an empty slot time. Let $P_t$ be the probability that the channel is busy in a slot because at least one station is transmitting. Then, it is $P_t = 1 - \prod_{s=1}^{N}(1-\tau^{(s)})$. The average idle slot duration can be defined as $T_I = P_I \cdot \sigma = (1 - P_t) \cdot \sigma$, where each idle slot is assumed to have duration $\sigma$.

**The average slot duration of a successful transmission.**
Consider a tagged station between the $N$ stations in the underlined network, and let $s$ be its index in the set $\{1, \ldots, N\}$. The probability that only the $s$-th tagged station is successfully transmitting over the channel can be defined as

$$P_s^{(s)} = \tau^{(s)} \prod_{\substack{j=1 \\ j \neq s}}^{N} (1 - \tau^{(j)}) \quad (13)$$

Then, the average slot duration of a successful transmission, which depends on the rate-class $(r)$ of the tagged station, can be evaluated as follows:

$$T_s^{(s)} = \frac{H_{PHY}}{R_C^{(r)}} + \frac{H_{MAC} + PL}{R_D^{(s)}} + \delta + \quad (14)$$
$$+ SIFS + \frac{H_{PHY} + ACK}{R_C} + \delta + DIFS$$

whereby $PL$ is the average payload length, $H_{PHY}$ and $H_{MAC}$ are, respectively, the physical and MAC header sizes, $\tau_p$ is the propagation delay, $DIFS$ is the duration of the Distributed InterFrame Space, $R_C$ is the basic data rate used for transmitting protocol data, and $R_D^{(s)}$ is the data rate of the $s$-th station.

With this setup, the average slot duration of a successful transmission can be evaluated as $T_S = \sum_{i=1}^{N} P_s^{(i)} \left(1 - P_e^{(i)}\right) \cdot T_s^{(i)}$.

**The average collision slot duration.** In a network of stations transmitting equal length packets with different data rates, the average duration $T_C$ of a collision is largely dominated by the slowest transmitting stations. This phenomenon is called *performance anomaly* of 802.11b, and it has been firstly observed in [12]. As an example, suppose that a frame transmitted by a station using the rate 1 Mbps (class 1) collides with the packet of a station transmitting at the bit rate 11 Mbps (class 4). Of course, both frames get lost while the channel appears as busy to the remaining sensing stations for the whole duration of the frame transmitted by the low rate station. Therefore, fast stations (higher classes) are penalized by the slow stations (low classes), causing a decrease of the throughput. In order to evaluate the collision probability, we define the class $(r)$ collision duration as $T_c^{(r)} = \frac{H_{PHY}}{R_C} + \frac{H_{MAC}+PL}{R_D^{(r)}} + ACK_{timeout}$, which takes into account the basic rate $R_C$ along with the data rate $R_D^{(r)}$ of the class $(r)$.

For the derivations which follow, we consider a set of indexes which identify the stations transmitting with the $r$-th data rate:

$n(r) = \{\text{identifiers of stations belonging to rate-class (r)}\}$

$\forall r \in R = \{1, \ldots, N_R\}$ such that $\sum_{r=1}^{N_R} |n^{(r)}| = N$ ($|\cdot|$ is the cardinality of the embraced set). With this setup, we notice two different collisions:

- intra-class collisions between at least two frames belonging to the same class rate $(r)$;
- inter-class collisions between *at least* one frame of class $(r)$ and *at least* one frame belonging to a class $(j) > (r)$

As far as intra-class $(r)$ collisions are concerned, the collision probability $P_{c1}^{(r)}$ can be evaluated as follows:

$$\left\{1 - \left[\prod_{s \in n(r)}(1-\tau^{(s)}) + \sum_{s \in n(r)} \tau^{(s)} \prod_{\substack{j \in n(r) \\ j \neq s}}(1-\tau^{(j)})\right]\right\} \cdot \prod_{s \in \{S-n(r)\}}(1-\tau^{(s)}) \quad (15)$$

Notice that the latter is the probability that the stations not belonging to the same data rate set $n(r)$, do not transmit, times the probability that there are at least two stations in the same rate class $n(r)$ transmitting over the channel. Notice that the first product within brace brackets accounts for the scenario in which the stations with rate in the set $n(r)$ are silent, or there is only a station transmitting with rate in the set $n(r)$. As a note aside, notice that $P_{c1}^{(r)} = 0$ if there are no collisions between stations belonging to the same rate class.

Following a similar reasoning, the inter-class $(r)$ collision probability $P_{c2}^{(r)}$ can be evaluated as:

$$\left[1 - \prod_{s \in n(r)}(1-\tau^{(s)})\right] \cdot \left[1 - \prod_{j=r+1}^{N_R} \prod_{s \in n(j)}(1-\tau^{(s)})\right] \cdot \left[\prod_{j=1}^{r-1} \prod_{s \in n(j)}(1-\tau^{(s)})\right] \quad (16)$$

which considers the scenario in which at least one station of class $(r)$ and at least one station belonging to a higher rate class (i.e., $(j) > (r)$) transmit in the same slot time, while all the other stations belonging to lower indexed classes (i.e., $(j) < (r)$) are silent. As a note aside, notice that $P_{c2}^{(r)} = 0$ if there are no collisions between stations belonging to different rate classes.

The total class $(r)$ collision probability is the sum of the previous two probabilities:

$$P_c^{(r)} = P_{c1}^{(r)} + P_{c2}^{(r)} \quad (17)$$

while the average collision slot duration can be computed considering the whole set of classes $r \in R$ along with their collision probabilities weighted by the respective durations:

$$T_C = \sum_{r=1}^{N_R} P_c^{(r)} \cdot T_c^{(r)} \quad (18)$$

**The average duration of the slot due to erroneous transmissions.** The average duration of the slot due to erroneous transmissions can be evaluated in a way similar to the one used for evaluating $T_S$ and $T_C$:

$$T_E = \sum_{i=1}^{N} P_s^{(i)} \cdot P_e^{(i)} \cdot T_e^{(i)} \quad (19)$$

whereby $P_s^{(i)}$ is defined in (13), and $T_e^{(s)}$ is assumed to be equal to $T_c^{(s)}$ since the transmitting station does not receive the acknowledgment before the end of the ACK timeout in the presence of channel errors.

**Average time slot duration.** Given the average slot durations derived in the previous sections, the average duration of a slot time can be evaluated as follows:

$$T_{av} = T_I + T_C + T_S + T_E \quad (20)$$

TABLE I
PHY SETUP OF THE IEEE 802.11B STANDARD

| Frequency [GHz] | 2.4 | 2.4 | 2.4 | 2.4 |
|---|---|---|---|---|
| Bit rate [Mbps] | 1 | 2 | 5.5 | 11 |
| Modulation | DBSPK | DQPSK | CCK | CCK |
| Chips per symbol, $C_s$ | 11 | 11 | 8 | 8 |
| Bits per symbol, $B_s$ | 1 | 2 | 4 | 8 |
| Channel band, $B_w$ [MHz] | 22 | 22 | 22 | 22 |
| Receiver Sensitivity AWGN-[dBm] | -85 | -82 | -80 | -76 |

TABLE II
TYPICAL NETWORK PARAMETERS

| MAC header | 28 bytes | Propag. delay $\tau_p$ | 1 $\mu s$ |
|---|---|---|---|
| PLCP Preamble | 144 bit | PLCP Header | 48 bit |
| PHY header | 24 bytes | Slot time | 20 $\mu s$ |
| basic rate | 1Mbps | $W_0$ | 32 |
| No. back-off stages, m | 5 | $W_{max}$ | 1024 |
| Payload size | 1028 bytes | SIFS | 10 $\mu s$ |
| ACK | 14 bytes | DIFS | 50 $\mu s$ |
| ACK timeout | 364$\mu s$ | EIFS | 364 $\mu s$ |

## V. TRAFFIC MODEL

This section presents the traffic model employed in our setup along with the derivation of the key probabilities $q^{(s)}$ and $P_{I,0}^{(s)}$ shown in Fig. 1. The offered load related to each station is characterized by the parameter $\lambda^{(s)}$ representing the rate at which packets arrive at the $s$-th station buffer from the upper layers, and measured in $pkt/s$. The time between two packet arrivals is defined as *interarrival time*, and its mean value is evaluated as $\frac{1}{\lambda^{(s)}}$. One of the most commonly used traffic models assumes that the packet arrival process follows a Poisson distribution. The resulting interarrival times are exponentially distributed.

In the proposed model shown in Fig. 1, we need a probability $q^{(s)}$ that indicates if there is at least one packet to be transmitted in the queue. Probability $q^{(s)}$ can be well approximated in a situation with small buffer size [7], [8], [9] through the following relation:

$$q^{(s)} = 1 - e^{-\lambda^{(s)} \cdot T_{av}} \quad (21)$$

where $T_{av}$ is the *expected time per slot*, useful to relate the state of the Markov chain with the actual time spent by a station in each state. Such a time has been derived in (20). Under the hypothesis of systems employing small queues, probabilities $q^{(s)}$ and $P_{I,0}^{(s)}$ can be approximated considering that the probability that at least one packet arrives in the queue at the end of a successful packet transmission is the same as having at least one packet arrival in an average time slot duration. As a result of this simple approximation, it is $q^{(s)} = P_{I,0}^{(s)}$. Upon remembering (5) and (9), $\tau^{(s)}$ in (10) can be evaluated as follows:

$$\tau^{(s)} = \frac{q^{(s)} \left(1 - P_{eq}^{(s)}\right)^{-1}}{q^{(s)}(\alpha^{(s)} - 1) + 1} \quad (22)$$

Though simple, this approximation proved to be quite effective for predicting the aggregate throughput through simulation.

## VI. PHYSICAL LAYER MODELLING

In a scenario with $N$ contending stations randomly distributed around a common AP, throughput performance depends on the channel conditions experienced by any station in the network. Consider a contending station at a distance $d$ from the AP. Given the one-sided noise power spectral density, $N_o = -174$ dBm$|_{T=273K}$, the received SNR can be evaluated as [16]:

$$SNR_{\text{dB}} = P(d)|_{\text{dBm}} - N_o - B_w|_{\text{dB}} - N_F \quad (23)$$

whereby $N_F$ is the receiver noise figure (10dB), while $P(d)|_{\text{dBm}}$, the power received at a distance $d$, corresponds to

$$P(d)|_{\text{dBm}} = P_{tx}|_{\text{dBm}} - PL_{\text{dB}} \quad (24)$$

Based on FCC regulations, in the 2.4GHz ISM band the transmitted power $P_{tx}|_{\text{dBm}}$ amount to 20 dBm, or, equivalently, 100 mW, while $PL_{\text{dB}}$ is the so-called path-loss [16]:

$$PL_{\text{dB}} = PL_o|_{\text{dB}} + 10 \cdot n_p \log_{10}\left(\frac{d}{d_0}\right)$$

whereby $PL_o|_{\text{dB}} = -10\log_{10}\left(\frac{G_t G_r \lambda^2}{(4\pi)^2 d_0^{n_p}}\right)$. The path-loss exponent, $n_p$, depends on the specific propagation environment, and it ranges from 2 (free space propagation) to 3.5-4 for non-line-of-sight propagation, or multi-path fast fading conditions, in indoor environments [16]. The SNR per transmitted bit, $\gamma$, is defined as:

$$\gamma|_{\text{dB}} = SNR_{\text{dB}} + 10\log_{10}\left(\frac{C_s}{B_s}\right) \quad (25)$$

whereby $C_s$ stands for chips per symbol, while $B_s$ is the number of bits per transmitted symbol. Both $C_s$ and $B_s$ are summarized in Table I. BER performance of the various transmitting modes of IEEE802.11b are shown in (26) for Rayleigh fading conditions [15], [17]:

$$\begin{array}{ll} \text{DBPSK} & \frac{1}{2(1+\gamma)} \\ \text{DQPSK} & \frac{1}{2}\left[1 - \sqrt{\frac{\gamma\frac{\sqrt{2}}{2}}{1+\gamma\frac{\sqrt{2}}{2}}}\right] \\ \text{CCK-5.5/11 Mbps} & \frac{2^{(\alpha-1)}}{2^\alpha - 1} \sum_{i=1}^{\alpha-1} \frac{(-1)^{i+1} C_i^{\alpha-1}}{1+i+i\cdot\gamma} \end{array} \quad (26)$$

whereby $\alpha = 4$ for 5.5 Mbps, and 8 for 11 Mbps, and $C_i^{\alpha-1} = \frac{(\alpha-1)!}{i! \cdot (\alpha-1-i)!}$.

## VII. SIMULATION RESULTS AND MODEL VALIDATION

This section focuses on simulation results for validating the theoretical models and derivations presented in the previous sections. We have developed a C++ simulator modelling the DCF protocol details in 802.11b for a specific number of independent transmitting stations. The simulator considers an Infrastructure BSS (Basic Service Set) with an AP and a certain number of fixed stations which communicates only with the AP. For the sake of simplicity, inside each station there are only three fundamental working levels: traffic model generator, MAC and PHY layers. Traffic is generated following the exponential distribution for the packet interarrival times. Moreover, the MAC layer is managed by a state machine which follows the main directives specified in the standard [1], namely waiting times (DIFS, SIFS, EIFS), post-backoff, backoff, basic and RTS/CTS access mode. The typical MAC layer parameters for IEEE802.11b noted in Table II [1] have been used for performance validation.

For conciseness, in this paper we present a set of results related to following scenarios A number of 9 contending stations are randomly placed along a circle of radius $R$, while the AP is placed at the center of the area. Upon employing Equ.s (23)-(25) with $n_p = 4$ (typical of heavy faded Rayleigh channel conditions), we have chosen a distance $R = 20$ m in such a way that the SNR per transmitted bit is above the minimum sensitivity, specified in Table I, relative to the bit rate 11 Mbps. Such stations are in saturated conditions and have PAR $\lambda = 8$ kpkt/s. The payload size, assumed to be common to all the transmitting stations, is equal to 1028 bytes. In this scenario, another station, in the following identified as the slow station, is placed at 4 different distances from the AP in such a way that transmission occurs with the four bit rates envisaged within the 802.11b protocol.

The theoretical aggregate throughput in this scenario is depicted in Fig. 2 as a function of the PAR of the slow station. Curves in both subplots have been parameterized with respect to the bit rate of the slow station. Simulated points are noted with cross-points over the respective theoretical curves. The upper curves refer to ideal channel conditions, i.e., $PER = 0$, while the lower subplot represents a scenario in which the packets transmitted by all the stations are affected by a $PER = 8 \cdot 10^{-2}$, which is the worst-case situation related to the minimum sensitivity [1]. Some considerations are in order. Both subplots show that the aggregate throughput is significantly lower than 11 Mbps even though all the stations transmit at the highest bit rate (continuous curve). Moreover, such a throughput reduces as a far as the PAR $\lambda_{slow}$ increases reaching saturation values strongly influenced by the rate of the slowest station. A comparative analysis of the set of curves depicted in both subplots reveal the throughput reduction due to the presence of channel induced errors.

## VIII. Conclusions

In this paper, we have presented a multi-dimensional Markovian state transition model characterizing the DCF behavior at the MAC layer of the IEEE802.11 series of standards by accounting for channel induced errors and multirate transmission typical of fading environments, under both non-saturated and saturated traffic conditions. The modelling allows taking into consideration the impact of channel contention in throughput analysis which is often not considered or it is considered in a static mode by using a mean contention period.

Theoretical derivations were supported by simulations.

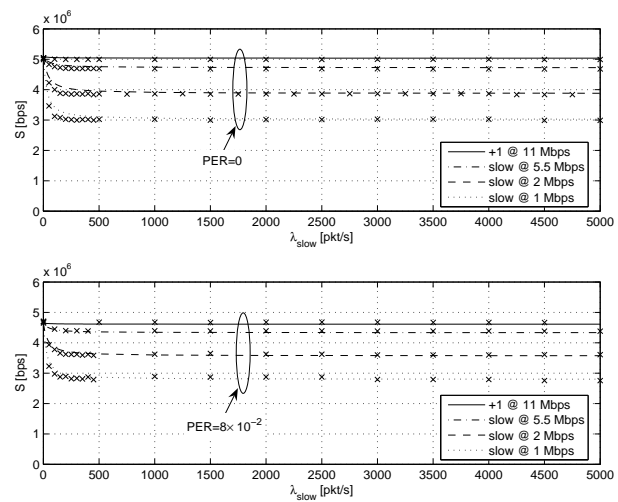

Fig. 2. Theoretical and simulated throughput for the 2-way mechanism as a function of the packet rate $\lambda_{slow}$ of the slow station, for four different bit rates, shown in the legends. Simulated points are identified by cross-markers over the respective theoretical curves.

## References

[1] *IEEE Standard for Wireless LAN Medium Access Control (MAC) and Physical Layer (PHY) Specifications*, November 1997, P802.11
[2] G. Bianchi, "Performance analysis of the IEEE 802.11 distributed coordination function", *IEEE JSAC*, Vol.18, No.3, March 2000.
[3] Ha Cheol Lee, "Impact of bit errors on the DCF throughput in wireless LAN over ricean fading channels", *In Proc. of IEEE ICDT '06*, 2006.
[4] Q. Ni, T. Li, T. Turletti, and Y. Xiao, "Saturation throughput analysis of error-prone 802.11 wireless networks", *Wiley Journal of Wireless Communications and Mobile Computing*, Vol. 5, No. 8, pp. 945-956, Dec. 2005.
[5] P. Chatzimisios, A.C. Boucouvalas, and V. Vitsas, "Influence of channel BER on IEEE 802.11 DCF", *IEE Electronics Letters*, Vol.39, No.23, pp.1687-1689, Nov. 2003.
[6] L. Yong Shyang, A. Dadej, and A.Jayasuriya, "Performance analysis of IEEE 802.11 DCF under limited load", *In Proc. of Asia-Pacific Conference on Communications*, Vol.1, pp.759 - 763, 03-05 Oct. 2005.
[7] D. Malone, K. Duffy, and D.J. Leith, "Modeling the 802.11 distributed coordination function in non-saturated heterogeneous conditions", *IEEE-ACM Trans. on Networking*, vol. 15, No. 1, pp. 159172, Feb. 2007.
[8] F. Daneshgaran, M. Laddomada, F. Mesiti, and M. Mondin, "Unsaturated Throughput Analysis of IEEE 802.11 in Presence of Non Ideal Transmission Channel and Capture Effects," *IEEE Trans. on Wireless Communications*, Vol. 7, No. 3, March 2008.
[9] F. Daneshgaran, M. Laddomada, F. Mesiti, and M. Mondin, "A Model of the IEEE 802.11 DCF in presence of non ideal transmission channel and capture effects," *In Proc. of IEEE Globecom 07*, Washington DC, November 2007.
[10] F. Daneshgaran, M. Laddomada, F. Mesiti, and M. Mondin, "On the linear behaviour of the throughput of IEEE 802.11 DCF in non-saturated conditions", *IEEE Communications Letters*, Vol. 11, No. 11, pp. 856-858, Nov. 2007.
[11] D. Qiao, S. Choi, and K.G. Shin "Goodput analysis and link adaptation for IEEE 802.11a wireless LANs", *IEEE Trans. On Mobile Computing*, Vol.1, No.4, Oct.-Dec. 2002.
[12] Heusse M.,Rousseau F., Berger-Sabbatel G. and Duda A. "Performance anomaly of 802.11b" *In Proc. of IEEE INFOCOM 2003*, pp. 836-843.
[13] D.-Y. Yang, T.-J. Lee, K. Jang, J.-B. Chang, and S. Choi, "Performance enhancement of multirate IEEE 802.11 WLANs with geographically scattered stations," *IEEE Trans. on Mobile Computing*, vol.5, no.7, pp.906-919, July 2006.
[14] G.R. Cantieni, Q. Ni, C. Barakat, and T. Turletti, "Performance analysis under finite load and improvements for multirate 802.11," *Computer Communications*, Elsevier, vol.28, pp.1095-1109, 2005.
[15] M.K. Simon and M. Alouini, *Digital Communication over Fading Channels: A Unified Approach to Performance Analysis*, Wiley-Interscience, 1st edition, 2000.
[16] T. S. Rappaport, *Wireless Communications, Principles and Practice*, Prentice-Hall, 2nd edition, USA, 2002.
[17] M. Fainberg, *A Performance analysis of the IEEE 802.11b local area network in the presence of bluetooth personal area network,* Available at http://eeweb.poly.edu/dgoodman/fainberg.pdf.